\documentclass[sigconf,natbib=true,nonacm]{acmart}
\AtBeginDocument{%
  }

\setcopyright{acmlicensed}
\copyrightyear{2018}
\acmYear{2018}
\acmDOI{XXXXXXX.XXXXXXX}
\acmConference[Conference acronym 'XX]{Make sure to enter the correct
  conference title from your rights confirmation email}{June 03--05,
  2018}{Woodstock, NY}
\acmISBN{978-1-4503-XXXX-X/2018/06}

\usepackage{caption}
\usepackage{subcaption}

\begin{document}

\title{Agent-centric Information Access}

\author{Evangelos Kanoulas, Panagiotis Eustratiadis, Yongkang Li, Yougang Lyu, Vaishali Pal, \\ Gabrielle Poerwawinata, Jingfen Qiao, Zihan Wang}
\affiliation{
  \institution{University of Amsterdam} 
  \country{The Netherlands}
}
\email{{ e.kanoulas, p.efstratiadis, y.li7, y.lyu, v.pal, g.poerwawinata, j.qiao, z.wang2 } @uva.nl}

\renewcommand{\shortauthors}{Kanoulas et al.}

\begin{abstract}
As large language models (LLMs) become more specialized, we envision a future where millions of expert LLMs exist, each trained on proprietary data and excelling in specific domains. In such a system, answering a query requires selecting a small subset of relevant models, querying them efficiently, and synthesizing their responses. This paper introduces a framework for agent-centric information access, where LLMs function as knowledge agents that are dynamically ranked and queried based on their demonstrated expertise. Unlike traditional document retrieval, this approach requires inferring expertise on the fly, rather than relying on static metadata or predefined model descriptions. This shift introduces several challenges, including efficient expert selection, cost-effective querying, response aggregation across multiple models, and robustness against adversarial manipulation. To address these issues, we propose a scalable evaluation framework that leverages retrieval-augmented generation and clustering techniques to construct and assess thousands of specialized models, with the potential to scale toward millions.

\end{abstract}

\maketitle

\section{Introduction}

\begin{figure*}[t]
        \centering
        \includegraphics[width=0.95\linewidth]{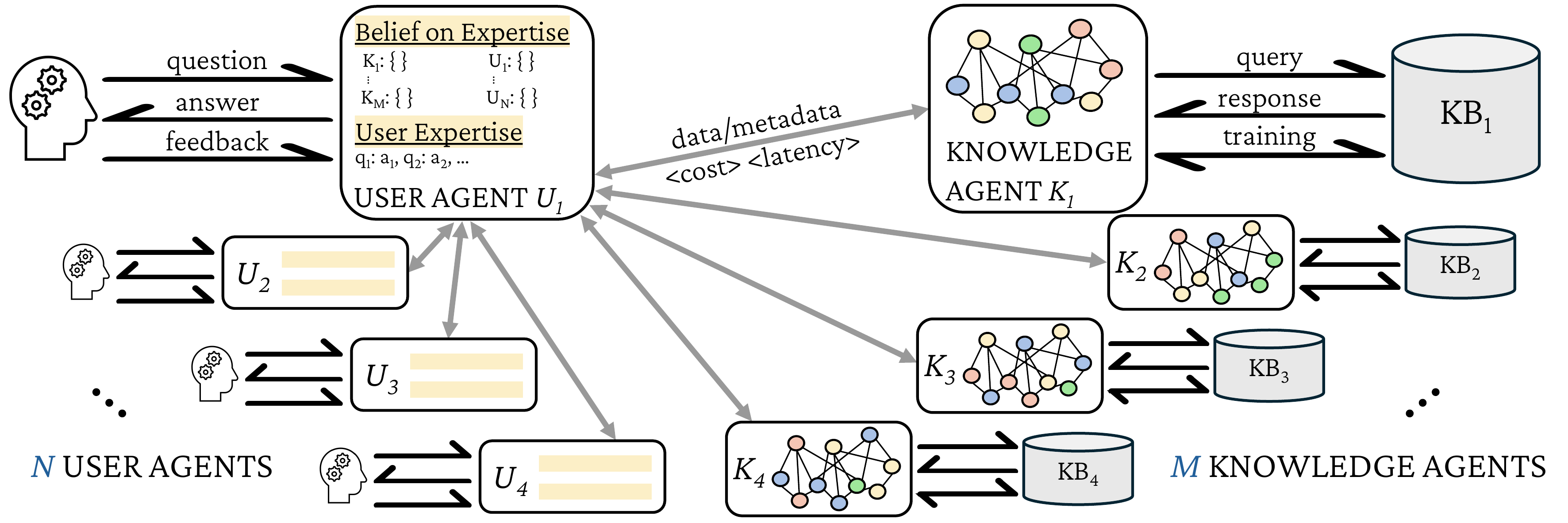}
        \caption{An illustration of a framework for querying both domain-specialized LLMs (knowledge agents) and user-specific LLMs (user agents). Users interact with the system through personalized user agents, which not only track past queries and responses to refine retrieval but also share specialized user knowledge with others. A belief model on expertise determines which knowledge agents ($K_1, K_2, \dots, K_M$) and user agents are experts on specific topics and should be queried to generate the most relevant answers. The system optimizes for cost and latency, ensuring efficient and accurate responses while minimizing unnecessary queries. This architecture enables the dynamic orchestration of expert LLMs, moving beyond document retrieval toward a multi-expert synthesis of knowledge.}
    \label{fig:metaLLM}
\end{figure*}

The rapid advancement of generative artificial intelligence has significantly reshaped the field of information retrieval (IR)~\cite{zhu2023large,allan-2024-future,white2025information}. Traditionally, a user-provided query prompts an IR system to align the query's intent with a pre-existing corpus of textual resources, returning a ranked list of documents from which the user then extracts relevant information. In this paper, we envision a future beyond traditional information access, in which information is served to users through large language models (LLMs)~\cite{openai2022chatgpt, wei2022emergentabilitieslargelanguage}. As generative AI models increasingly act as primary sources of information, this shift necessitates a new paradigm. This paradigm must address the challenges of selecting the most suitable LLMs to query and synthesizing knowledge across multiple expert models.

This future is unfolding at a fast pace~\cite{caramancion2024large}\footnote{\url{https://kafkai.com/en/blog/ai-impact-on-search-engines/}}, driven by the proliferation of domain-specialized LLMs tailored to increasingly narrow fields of expertise. We are moving toward a world where millions of such expert LLMs will exist, each trained on distinct knowledge domains, some overlapping, others highly specialized. This shift is already underway, with companies like BloombergGPT~\cite{wu2023bloomberggpt}, Thomson Reuters\footnote{\url{https://legalsolutions.thomsonreuters.co.uk/en/c/practical-law/now-with-generative-ai.html}}, Elsevier\footnote{\url{https://www.elsevier.com/products/scopus/scopus-ai}} developing LLMs on proprietary material. This trend is set to accelerate as more content providers, institutions, and organizations build models fine-tuned to their unique datasets.
As this paradigm evolves, a user's information needs will increasingly be met not by a single system, but by querying and synthesizing responses from multiple expert LLMs, requiring careful orchestration to extract the most accurate and relevant insights.

Beyond expert LLMs, we envision a future where users themselves become ``experts'' through their search behavior. Personalized digital assistant LLMs will continuously read, process, and synthesize everything a user has searched for and engaged with, leading to the creation of millions of individualized user-LLMs, each evolving to reflect a specific user's knowledge, expertise, and interests.

In this evolving digital ecosystem, an agent will be necessary to orchestrate queries and synthesize responses from domain-expert LLMs. However, querying all available LLMs is prohibitively expensive and inefficient. Thus, a new type of search engine is required that, given a user's question, intelligently retrieves only the most relevant expert LLMs. This agent must dynamically determine which expert models to consult, balancing the need for accuracy with cost efficiency. Upon receiving a user's query, it should strategically decide which LLMs to engage, ensuring the response is both comprehensive and cost-effective while minimizing redundant computations. 

This new paradigm presents several fundamental challenges. First, selecting relevant LLM experts by querying all available models is computationally expensive and inefficient, requiring intelligent mechanisms to identify the most relevant models based on expertise, context, and cost. Second, multi-LLM answer aggregation introduces new complexities, as responses from different expert LLMs must be synthesized, reconciled, and ranked to provide coherent and contextually accurate answers. Third, bias, trust, and adversarial manipulation in multi-LLM retrieval become pressing concerns, as LLM-generated responses may reflect underlying biases, selectively omit critical information, or be manipulated by adversarial actors. Finally, these challenges require a novel evaluation framework and the infrastructure necessary to study. Addressing these issues will be crucial in shaping the next generation of model-centric\footnote{In this work we use model-centric and LLM-centric interchangeably.} IR.

To address these challenges, we propose a framework for orchestrating domain-specialized LLMs (knowledge agents) and personalized user LLMs (user agents). As illustrated in Figure \ref{fig:metaLLM}, this framework structures IR as a dynamic interaction between users and a distributed network of expert models.
Users interact with the system through personalized user agents, which refine retrieval strategies based on past queries, responses, and feedback. These agents maintain a belief model on expertise, dynamically assessing which knowledge agents (domain-specialized LLMs) and user agents (peers with relevant expertise) are most suitable for responding to a given query. This belief model is continuously updated based on user interactions, ensuring that information synthesis prioritizes the most reliable and contextually relevant sources.
Each knowledge agent specializes in a distinct domain, accessing structured knowledge bases (KBs) or proprietary datasets. Some knowledge agents overlap in expertise, while others are highly specialized, creating a network of interoperable expert models that can be selectively queried based on the belief model's assessment. Additionally, user agents share specialized user knowledge, allowing insights from personalized experiences to propagate through the system.
To optimize cost and latency, the framework incorporates an adaptive query mechanism, reducing redundant queries and minimizing computational overhead while maintaining response quality. Instead of blindly querying all potential sources, the system selectively engages only the most relevant knowledge and user agents, balancing efficiency with accuracy.

To ground our discussion, we first examine previous and current IR paradigms relevant to the  multi-LLM retrieval paradigm. We then explore each of the key challenges in detail, analyzing the complexities of selecting relevant LLM experts, aggregating responses across multiple models, ensuring trust and robustness. Finally, we introduce a simplified evaluation framework for ranking relevant LLM experts as a starting point.
\section{Related Work}

In this section we focus on two topics that bear similarities to the proposed model-centric information access paradigm, (i) previous retrieval paradigms in which multiple collections/models were combined to produce the final ranking of resources upon a user's query, namely distributed and federated retrieval and meta-search, and (ii) Agentic AI as an emerging research topic where AI agents plan the solutions to complex problems engaging a large number of specialized expert models in the execution of their plans.

\subsection{Distributed Information Retrieval, Federated Search, and Meta-Search}

Multi-LLM retrieval shares conceptual similarities with distributed information retrieval (DIR)~\cite{Callan1995, Callan2000, Crestani2013}, federated search (FS)~\cite{Shokouhi2009}, and meta-search~\cite{Glover2000metasearch, Chen2001Metasearch}, as all involve querying and integrating results from multiple independent sources. However, while these traditional approaches rely on structured document collections with predefined ranking mechanisms, multi-LLM retrieval must dynamically infer and assess expertise, making it a fundamentally different and more complex problem.

Meta-search engines issue queries to multiple independent search engines and aggregate their results using fusion techniques such as reciprocal rank weighting or probabilistic rank merging~\cite{Glover2000metasearch}. Similarly, federated search queries multiple independent databases and merges the retrieved results into a unified ranking. However, unlike these methods, which rely on external document rankings, expert LLM retrieval must actively determine which models to engage first and then synthesize their responses. Moreover, meta-search systems depend on external ranking functions provided by the underlying search engines, while multi-LLM retrieval lacks predefined ranking mechanisms, requiring alternative strategies such as confidence-based scoring, response quality estimation, and adaptive model selection. Additionally, whereas meta-search/federated search operates on pre-indexed documents, multi-LLM retrieval generates responses dynamically, introducing further challenges in answer aggregation, conflict resolution, and provenance tracking~\cite{Glover2001metasearcharchitecture}. Finally, result merging in meta-search typically involves normalizing and fusing ranked lists, whereas multi-LLM retrieval synthesizes knowledge across models, demanding approaches beyond traditional rank fusion~\cite{Shokouhi2009}.

DIR shares similarities with multi-LLM retrieval in that queries are processed across multiple autonomous document collections, with ranking determined by each collection's estimated ability to satisfy the information need~\cite{Callan1995}. The DIR process consists of three key steps: (i) \textit{resource description}, where each collection is characterized using high-level metadata; (ii) \textit{resource selection}, where a subset of collections is chosen based on estimated query relevance~\cite{Callan1995, Callan2000}; and (iii) \textit{results merging}, where retrieved documents from selected sources are integrated into a unified ranked list~\cite{Crestani2013}.

Despite these similarities, a key distinction is that DIR relies on pre-existing document collections, where relevance can be inferred through established statistical measures. In contrast, multi-LLM retrieval does not operate over a fixed set of indexed documents but rather dynamically selects and queries LLMs that generate responses on demand. Unlike DIR, where source selection is driven by stable collection statistics, multi-LLM retrieval must infer expertise dynamically, since LLMs lack structured metadata and predefined ranking signals. This lack of explicit document collections and structured retrieval units makes expertise assessment and reliability estimation significantly more challenging in multi-LLM environments.

\subsection{Agentic AI and Multi-Agent Collaboration}

An agentic information access ecosystem also bears similarities with agentic AI~\cite{acharya2025agentic}. Agentic AI is a fast-progressing field where AI systems are composed of autonomous, interacting agents that can plan, adapt, and collaborate to achieve complex goals with minimal human intervention. These agents can independently perceive environments, make decisions, and take actions, often coordinating with other agents or tools to optimize outcomes. Key challenges in agentic AI include coordination and orchestration, ensuring that multiple agents effectively divide tasks and avoid conflicts; expertise inference, dynamically selecting the best agent for a given task without predefined metadata; trust and robustness, preventing biases, adversarial manipulation, or errors from propagating across the system; scalability, as increasing the number of agents introduces computational and communication bottlenecks; and evaluation, since traditional AI benchmarks do not capture the complexities of multi-agent interactions, making it difficult to assess decision quality, efficiency, and emergent behaviors.

Despite these overlaps, the primary focus of agentic AI is on planning and execution, where AI agents strategically break down tasks, select relevant tools or models, and iteratively refine their outputs~\cite{li2023camelcommunicativeagentsmind}. In contrast, agentic information access focuses on retrieving and synthesizing knowledge, where the challenge lies not in executing multi-step plans but in identifying the most relevant information sources and integrating their responses effectively.

Second, most agentic AI systems are constrained to a limited number of agents, typically orchestrating a handful of models or API calls to accomplish a goal~\cite{hong2024metagptmetaprogrammingmultiagent, wu2023autogenenablingnextgenllm}. In contrast, agentic information access envisions a large-scale information ecosystem, where potentially thousands or even millions of domain-specialized LLMs must be efficiently queried, ranked, and synthesized. This distinction raises scalability challenges that agentic AI frameworks have yet to address, such as query budget optimization, redundancy elimination, and cost-aware model selection.

A third fundamental difference lies in the availability and structure of metadata. In current agentic AI frameworks, each tool, API, or model is associated with structured or semi-structured metadata, such as a description of its capabilities, and supported inputs/outputs~\cite{xu2023toolbench, li2023apibank}. This metadata provides a prior for model selection, simplifying the decision-making process. For instance, Hugging Face model cards ~\cite{Mitchell2019}\footnote{\url{https://huggingface.co/docs/hub/en/model-cards}} provide descriptive summaries that allow an AI agent to infer whether a model is suited for a given task. However, in multi-LLM retrieval, no standardized metadata can be assumed for all models, making expertise identification significantly harder. Unlike APIs that perform heterogeneous tasks -- such as image segmentation, text-to-speech conversion, or protein folding -- the LLMs in a multi-LLM retrieval system are more homogeneous, all designed to generate textual responses. This lack of clear differentiation complicates expert selection, as distinguishing models requires indirect signals, such as previous retrieval effectiveness, self-reported confidence, or trust-based reputation scores.
\section{Challenges in Multi-LLM Information Retrieval}
\label{sec:challenges}

The transition from document-centric to model-centric IR introduces a range of fundamental challenges. Unlike conventional search engines that retrieve and rank documents based on well-established data structures and ranking functions, an LLM-based system must dynamically select, query, and synthesize responses from a diverse set of expert models. This raises critical questions:
\begin{itemize}
    \item[RQ1] Which LLMs should be queried for a given information need? 
    \item[RQ2] How should their responses be aggregated into a coherent answer? 
    \item[RQ3] How do we ensure robustness against bias or adversarial manipulation? 
    \item[RQ4] How do we evaluate a model-centric retrieval system?
\end{itemize}

In this section, we explore these challenges in detail, starting with the problem of selecting relevant LLM experts, followed by multi-LLM answer aggregation, issues of robustness, and the development of novel evaluation frameworks.

\subsection{Selecting Relevant LLM Experts}
Selecting the right LLMs for a given query is a fundamental challenge. Unlike traditional IR, where search engines rely on well-established ranking functions over documents, model-centric search requires identifying and querying a subset of expert models, balancing accuracy, diversity, cost, and efficiency.

\subsubsection{Defining LLM Expertise}

A fundamental challenge in an LLM-centric retrieval system is determining what makes an LLM an expert on a given topic. Unlike traditional document retrieval, where relevance can be inferred from topical similarity measures, expertise in LLMs is more elusive. Should an LLM be considered an expert based on the data it was trained on, the domains it was fine-tuned for, or its demonstrated ability to provide high-quality answers? In many cases, LLMs operate as black boxes, offering little transparency about their training data or the methods by which they generate responses. This lack of visibility complicates the process of verifying their domain competence. Additionally, even if an LLM is trained on relevant material, does that necessarily mean it can generate accurate, contextually appropriate responses? Expertise is not just about knowledge but also about reasoning and the ability to contextualize information correctly. Given this, we need a systematic way to rank LLM expertise dynamically, rather than relying on static, pre-determined categorizations. This leads to the following research question:

\begin{itemize}
    \item[\textbf{RQ1.1}] How can we define and measure the expertise of an LLM?
\end{itemize}

One possible approach is to infer expertise empirically by evaluating an LLM's responses on benchmark datasets covering different domains. However, this method is limited to the domains covered by the benchmark. Further, there is no guarantee that all LLMs can be evaluated against the same benchmarks. Alternatively, reputation-based models could be developed, where an LLM's past performance on user queries informs its future ranking for similar queries. But this raises additional concerns: how do we assess the performance of an LLM? Do we need human judges, automatic methods to assess performance, e.g. new query performance prediction methods, or new uncertainty quantification methods that correlate well with accuracy? How do we handle new or emerging LLMs that lack historical performance data? And how can we ensure that performance metrics are robust against adversarial manipulation, where LLM providers game the system to appear more authoritative than they actually are? 
Addressing these challenges requires a fundamental rethink of how expertise is defined, measured, and updated in a rapidly evolving LLM ecosystem.

\subsubsection{Efficient Querying}

Once relevant LLM experts have been identified, the next challenge is determining how many should be queried to maximize response quality while minimizing cost and computational overhead. Unlike traditional search engines, where retrieving more documents often improves recall without significant additional expense, querying more LLMs incurs non-trivial costs in computation, latency, and financial resources. Each LLM query consumes processing power and may be subject to usage fees, making an exhaustive search across all available experts infeasible. Instead, an efficient expert selection strategy must strike a balance: querying too few LLMs may result in incomplete or biased answers, while querying too many can lead to unnecessary redundancy, slower response times, and increased costs. This leads to the following research question:

\begin{itemize}
    \item [\textbf{RQ1.2}] How many LLMs should be queried once we identify a ranking of expertise for a given query?
\end{itemize}

An ideal system would dynamically adjust the number of experts consulted based on the complexity of the query and the expected information gain from additional models. If a simple factual question can be confidently answered by a single high-reputation LLM, there is little benefit in polling multiple experts. However, for ambiguous, multi-faceted, or domain-spanning queries, aggregating responses from multiple LLMs might be necessary to construct a well-rounded answer. The system must therefore develop predictive mechanisms to estimate, before execution, how much additional querying will improve the response. 

\subsubsection{Cost and Query Budgeting: How Do We Minimize Computational Overhead?}

Another consideration is how to prioritize and schedule LLM queries. Should the system query all selected LLMs in parallel to minimize latency, or should it use an adaptive querying strategy, where it starts with a small number of high-confidence experts and only expands the search if the responses are inconsistent or incomplete? 

\begin{itemize}
    \item [\textbf{RQ1.3}] How can we minimize the computational overhead of querying LLMs?
\end{itemize}

A practical approach is to implement a tiered querying strategy based on cost and expected value. Low-cost or freely available LLMs could be queried first, providing an initial response, with more expensive models accessed only if additional expertise is needed. This strategy is analogous to progressive search techniques in IR, where retrieval starts with a small, efficient subset before expanding if necessary. However, to implement such a system effectively, the system must be able to predict the expected cost and utility of querying different LLMs in real time. This requires building cost-aware meta-indices, which estimate an LLM's likely expertise, confidence, and query cost before actually invoking it.

A more adaptive solution would involve reinforcement learning or multi-armed bandit approaches that dynamically optimize the cost-performance trade-off. The system could learn over time which LLMs provide the most useful answers for specific query types, gradually refining its querying strategy to minimize cost without sacrificing accuracy. Additionally, cost modeling could incorporate user preferences and query criticality -- for instance, users might accept slightly less accurate responses for routine queries but demand higher-quality results for complex, high-stakes questions, justifying higher costs.

Ultimately, achieving cost-effective querying in an LLM-based IR system requires a multi-layered optimization approach: (i) predictive cost modeling to estimate the utility of each LLM before querying it, (ii) adaptive, tiered querying strategies that escalate complexity only when needed, and (iii) learning-based approaches to refine expert selection dynamically. Without such mechanisms, the financial and computational burden of large-scale LLM-based retrieval would be prohibitive, limiting the viability of this new paradigm.\looseness=-1

\subsubsection{Temporal and Contextual Adaptation: Can LLM Expertise Change Over Time?}
Expertise is not static. What is considered authoritative today may become outdated tomorrow. In an LLM-centric retrieval system, this challenge is even more pronounced because LLMs do not inherently update their knowledge in real time. Unlike traditional search engines, which can index new documents as they become available, LLMs rely on their training data, which may be months or even years old. This creates a fundamental issue: how do we ensure that LLMs selected as experts remain relevant as real-world facts change? If an LLM trained on medical knowledge from 2022 is queried about a treatment breakthrough from 2024, its response may be inaccurate, yet the system may still consider it an "expert" based on its past performance. This leads to the following research question:

\begin{itemize}
    \item [\textbf{RQ1.4}] How can we update the expertise of LLMs in a dynamic fashion?
\end{itemize}

One approach to mitigating this problem is periodic re-evaluation of LLM expertise. Just as human experts require continuous education to stay updated, LLMs could be assessed on evolving benchmark datasets that incorporate the latest knowledge. However, this requires an infrastructure capable of continuously testing LLMs against new ground-truth information, a challenging and resource-intensive goal. 

\subsection{Multi-LLM Answer Aggregation}
Once multiple LLM experts are selected and queried, the next challenge is how to synthesize their responses into a coherent, accurate, and useful answer. Unlike traditional search, where ranked lists of documents allow users to extract information themselves, an LLM-centric retrieval system must integrate multiple expert outputs -- some of which may complement each other, while others may conflict. Below are the key open subtopics in multi-LLM answer aggregation.

\subsubsection{Response Merging}

A naive approach might be to concatenate all responses, but this often results in redundancy, verbosity, or inconsistencies. Instead, should aggregation rely on abstractive summarization techniques, where responses are fused into a more concise and cohesive format? Alternatively, should users be presented with a ranked list of expert opinions, preserving individual perspectives? Another consideration is context-dependent synthesis: should the method of merging depend on the type of question asked, with factual queries receiving a single integrated response while opinion-based or exploratory queries maintain multiple viewpoints? The challenge, then, is to design aggregation methods that minimize redundancy while ensuring completeness, clarity, and informativeness. This leads to the following research question:

\begin{itemize}
    \item [\textbf{RQ2.1}] How can we develop methods that optimally aggregate different answers account for the query and user needs?
\end{itemize}

\subsubsection{Conflict Resolution}
As different LLM experts are consulted, it is inevitable that some will provide conflicting answers, especially on ambiguous or evolving topics. Unlike traditional search, where users can compare multiple sources themselves, LLM aggregation systems must resolve inconsistencies automatically or present contradictions transparently. Should models be ranked based on confidence scores, with higher-confidence responses prioritized in synthesis? Should conflicting responses trigger a secondary verification process, where additional LLMs are queried as tie-breakers? Moreover, does presenting multiple perspectives side by side, rather than forcing a single consensus, better align with the needs of users searching for nuanced or controversial topics? The system must strike a balance: when should conflicts be merged into a neutral response, and when should disagreements be explicitly surfaced? Without robust mechanisms for handling contradictions, users may receive misleading or oversimplified information, undermining trust in the system. This leads to the following research question:

\begin{itemize}
    \item[\textbf{RQ2.2}] How do we present conflicting answers to users?
\end{itemize}

\subsubsection{Confidence Estimation}
Trustworthiness is a key concern in multi-LLM aggregation, as users need a way to assess the reliability of synthesized responses. Since different LLMs vary in training data, reasoning capabilities, and update frequency, some responses may be significantly more reliable than others. How should the system evaluate an LLM's confidence in its own response? Should models be required to self-report confidence scores, or should an external mechanism infer reliability? Could meta-evaluation models analyze linguistic patterns, factual consistency, or source attribution to predict response trustworthiness? Additionally, if multiple LLMs agree on an answer, should their collective agreement serve as an implicit confidence boost, akin to ensemble methods in machine learning? A key research challenge is to develop a framework where users can gauge not just what is being said, but how reliable the information is, ensuring that low-confidence or speculative responses are not treated as authoritative. This leads to the following research question:

\begin{itemize}
    \item[\textbf{RQ2.3}] How do we quantify the reliability of an answer in a way that we can compare multiple heterogeneous models?
\end{itemize}

\subsubsection{Attribution and Source Transparency}
One of the fundamental challenges in aggregating responses is ensuring transparency regarding where information comes from. Users of traditional search engines can inspect the provenance of retrieved documents, but in an LLM-centric retrieval system, the origin of synthesized information is often obscured. Should responses include explicit citations to the LLMs that contributed to the final answer? If so, how should attribution be handled when multiple models generate similar but slightly different responses? Furthermore, proprietary LLMs may refuse to disclose their sources, making it difficult to assess their credibility. Would a trust registry for LLMs, where models are scored based on their past citations and accuracy, be a viable solution? Transparency in attribution is not just about trust -- it also ensures accountability, preventing the spread of unverifiable or misleading information. Without clear provenance, users may struggle to distinguish authoritative knowledge from speculative or hallucinated responses. This leads to the following research question:

\begin{itemize}
    \item[\textbf{RQ2.4}] How do we attribute answers to the underlying models that provided them?
\end{itemize}

\subsection{Bias, Trust, and Adversarial Manipulation in Multi-LLM Retrieval}
In an LLM-centric retrieval system, trust and bias mitigation are fundamental challenges, especially when adversarial actors seek to manipulate perceived expertise for strategic gain. Unlike traditional search engines, where documents and sources can be manually verified, an LLM-based ecosystem relies on dynamic model selection, which opens the door for adversarial exploitation. Malicious entities could develop LLMs that appear authoritative, only to later inject biased, misleading, or harmful responses into the information ecosystem. This presents a novel and pressing research challenge: how can we ensure that expert selection remains unbiased, resistant to manipulation, and aligned with factual accuracy?

\subsubsection{Preventing the Domination of Certain LLMs in Expert Selection}
A major risk in multi-LLM retrieval is the monopolization of expert selection by a small set of dominant models. If the aggregation system frequently favors the same LLMs, users may receive an overly narrow perspective, reinforcing existing biases. Further, if user feedback is incorporated into the belief model for expertise selection, these biases can be amplified over time, undermining the quality and diversity of ranking. This raises the question: 

\begin{itemize}
    \item [\textbf{RQ3.1}] How do we ensure that expert selection is diverse and representative?
\end{itemize} 

One approach is to introduce diversification constraints, ensuring that responses incorporate perspectives from multiple LLMs, even if some models have lower confidence scores. Alternatively, should the system cap the frequency with which a specific LLM is selected per query type to prevent overrepresentation? However, these solutions introduce trade-offs: if an LLM is truly the most reliable expert, artificially limiting its influence could reduce overall answer quality. Another concern is gaming of the expert ranking system -- should model providers be able to pay for increased visibility, and if so, how do we prevent economic incentives from skewing expert selection? Without safeguards, an open-market approach could lead to a winner-takes-all scenario, where the most financially backed LLMs dominate, drowning out smaller but potentially more accurate models.

\subsubsection{Adversarial Attack Models}
In an open ecosystem where LLMs compete to be recognized as experts, malicious actors could design adversarial models that exploit trust mechanisms. 
A deceptive LLM might start by providing reliable, high-quality answers, gradually building a strong reputation within the system. However, once it achieves expert status, it could subtly introduce misinformation, steering responses in ways that serve political, economic, or ideological interests. Another potential attack vector is citation gaming, where multiple adversarial LLMs artificially reinforce each other's credibility by citing one another in their responses. This leads to the following research question:

\begin{itemize}
    \item [\textbf{RQ3.2}] What is an effective adversarial attack in a model-centric retrieval?
\end{itemize}

\subsubsection{Adversarial Attack Detection}

Detecting such adversarial behavior is an urgent challenge. One possible defense is delayed adversarial detection, where models are continuously evaluated over time for sudden shifts in behavior. Additionally, should trust scores decay over time if inconsistencies or contradictions are detected? Addressing these risks requires robust adversarial detection systems, perhaps borrowing techniques from fraud detection, anomaly detection, and adversarial machine learning. Without proactive safeguards, LLM-driven search could become an easy target for misinformation campaigns, with major societal implications. This leads to the following research question:

\begin{itemize}
    \item [\textbf{RQ3.3}] How can we robustify ranking algorithms against adversarial attacks?
\end{itemize}

To maintain trust in multi-LLM retrieval, a reputation framework must be developed to evaluate expert reliability over time. But what metrics should determine trust? Should trust scores be assigned based on historical accuracy? If so, how do we fairly measure accuracy in subjective domains like law, ethics, or politics? Another approach could involve user feedback, where responses are rated for correctness and usefulness. However, how do we prevent coordinated attacks, where adversarial actors manipulate ratings to boost or suppress certain LLMs? 

\subsection{Developing an Evaluation Framework for Multi-LLM Retrieval}

The evaluation of IR systems has traditionally relied on test collections consisting of queries, a fixed document corpus, and relevance judgments~\cite{Voorhees2001TREC, Robertson2008evaluation}. This structured approach has enabled the development of standard retrieval metrics, which assess how well a system retrieves and ranks documents. However, multi-LLM retrieval fundamentally alters the nature of retrieval units, shifting from static documents to generated responses that synthesize information from multiple models. This shift necessitates a re-examination of evaluation frameworks, as traditional IR methodologies do not fully capture the challenges introduced by generative responses, expertise selection, and response aggregation.

\subsubsection{Computational Challenges in Developing a Million LLMs}
Building a multi-LLM retrieval ecosystem at the scale of millions of specialized models introduces significant computational challenges that go beyond traditional machine learning infrastructures. Unlike current AI deployment paradigms, where a small number of general-purpose LLMs are trained and fine-tuned on diverse datasets, a system with millions of domain-specific LLMs requires an entirely new approach to training, and deployment.

A fundamental challenge lies in model differentiation and specialization. Training millions of LLMs with distinct expertise requires efficient knowledge partitioning, ensuring that each model acquires specialized domain knowledge while avoiding unnecessary redundancy. Traditional fine-tuning strategies, which involve adapting a base model to a specific dataset, do not scale well when training millions of models, as each instance would demand separate computational resources, training data pipelines, and continuous updates. Instead, novel parameter-efficient tuning techniques such as Low-Rank Adaptation (LoRA)\cite{Hu2021LoRA} and Mixture of Experts (MoE) architectures\cite{Shazeer2017MoE} may offer scalable alternatives by allowing models to share a common backbone while activating only domain-specific components during inference. A different scalable solution could involve architectures that integrate retrieval-augmented generation (RAG)~\cite{Lewis2020RAG} to allow LLMs to fetch information from a specialized data store. The hardware and storage infrastructure required to support a million LLMs is another critical bottleneck. Unlike traditional search engines, where documents are stored in an index, a multi-LLM system requires active model storage, retrieval, and execution.  This leads to the following research question:

\begin{itemize}
    \item [\textbf{RQ4.1}] How can we build the necessary infrastructure to support a large number of specialized LLMs?
\end{itemize}

\subsubsection{Redefining Test Collections for Multi-LLM Retrieval}
In docu-\\-ment-centric IR, test collections are built around a static corpus, allowing queries to be evaluated based on a predefined set of relevance judgments. In contrast, multi-LLM retrieval does not involve retrieving pre-existing documents but instead generates responses dynamically. This shift raises fundamental questions about how to construct evaluation datasets. Should test collections contain human-written reference responses, similar to QA and summarization benchmarks like SQuAD, MS MARCO, or Natural Questions~\cite{Rajpurkar2016SQuAD, Bajaj2018MSMARCO}? Alternatively, should multi-LLM retrieval be evaluated based on human comparative judgments, where different model outputs are rated and ranked? This leads to the following research question:\looseness=-1

\begin{itemize}
    \item [\textbf{RQ4.2}] How can we build static and reusable test collections to assess the LLM expertise and the quality of the response aggregation?
\end{itemize}

\subsubsection{New Metrics for Evaluating Multi-LLM Responses}
The transition from document ranking to generative response synthesis requires new evaluation metrics that go beyond traditional IR effectiveness measures. Precision, recall, and NDCG assume a fixed relevance scale based on pre-existing documents, making them unsuitable for evaluating generated responses that may vary significantly in wording and structure. One possible direction is to adapt existing NLP evaluation metrics such as BLEU~\cite{Papineni2002BLEU}, ROUGE~\cite{Lin2004ROUGE}, and BERTScore~\cite{Zhang2019BERTScore} to assess content overlap between generated and reference responses. However, these metrics primarily measure lexical or embedding similarity, with already known issues in generative model evaluation. This leads to the following research question:\looseness=-1

\begin{itemize}
    \item[\textbf{RQ4.3}] What metrics allow us to compare different user agents that search, retrieve, query and aggregate answers against multiple LLMs? 
\end{itemize}

\section{An Evaluation Framework for Multi-LLM Ranking}

In this section, we introduce a framework for evaluating agents in their ability to rank multiple LLMs based on their expertise. Our approach simplifies the broader problem of generating optimal answers while minimizing query costs, allowing us to focus specifically on the ranking aspect. These simplifications enable the construction of a reusable and standardized evaluation infrastructure.\looseness=-1

\subsection{Problem Definition}

Let $l \in L$ represent an LLM expert model within a collection of expert LLMs, $L$. Given a user query $q$, the ultimate objective is to generate an optimal answer $a$ while minimizing the computational and financial costs of querying models in $L$.

To facilitate a controlled evaluation, we decompose the problem into two stages: (i) expertise identification and (ii) answer generation. The former is about determining which LLMs are most relevant to a query before invoking them for answer generation, while the latter concerns selecting and querying the most relevant LLMs to produce a response. To isolate the ranking problem, we simplify the task as follows:  
\begin{itemize}
    \item Upon receiving a user query $q$, the user agent must rank all LLMs based on their belief about which LLMs are most relevant to answering $q$.
    \item Prior to real-time querying, the agent is provided with a predefined budget $b$, allowing it to query a subset of LLMs across a set of preparatory queries $\{q\}$.
    \item The challenge is to develop a strategy that maximizes information acquisition about each LLM while minimizing the number of queries issued, ultimately leading to better ranking accuracy for an unseen query $q'$.
\end{itemize}

When an agent queries an LLM $l$ with a query $q$, the model returns:  
\begin{itemize}
    \item An answer $a$,
    \item Metadata $m$, such as information about the model (e.g., ``model cards''), probabilities estimated during auto-regressive generation (used to calculate metrics like perplexity\footnote{\url{https://huggingface.co/spaces/evaluate-metric/perplexity}}~\cite{jelinek1977perplexity}~), logits from the final neural network layer (which could be leveraged for uncertainty estimation), and others.
\end{itemize}

This formulation remains a simplification of the full problem, which ideally would integrate expertise identification with answer generation -- allowing the agent to simultaneously answer user queries while updating its belief on LLM expertise, and allow unrestricted querying during expertise identification -- permitting the agent to use arbitrary query strategies rather than a predefined set of queries. However, by constraining the problem in this way, we enable the construction of a reusable test collection, consisting of standardized query-response metadata.

\subsection{Reusable Test Collection for Evaluation}
\label{sec:test_collection}
A well-structured test collection facilitates systematic evaluation and reproducibility. We propose structuring the dataset as follows:

\subsubsection{Training Data}  

A training set provides access to LLM responses for a range of preparatory queries, allowing agents to learn expertise patterns. Each training tuple consists of:
\[
(q_i, a_i^j, m_i^j)
\]
where $q_i$ is a user query, $a_i^j$ is the answer from LLM $l_j$, $m_i^j$ represents metadata provided by $l_j$, such as confidence scores, logits, or retrieval information. This data allows ranking models to develop heuristics from past responses and model-specific characteristics.

\subsubsection{Test Data}

The test set evaluates the agent's ability to rank LLMs effectively. It can be structured in two ways:

\begin{enumerate}
    \item Answer-Based Evaluation: Comparing LLM-generated answers against ground-truth responses:
    \[
    (q_i, a_i^G)
    \]
    where $a_i^G$ is the correct reference answer for query $q_i$.

    \item Relevance-Based Evaluation: Measuring the relevance of each LLM for a given query:
    \[
    (q_i, qrels_i)
    \]
    where $qrels_i$ defines the relevance scores of all LLMs for query $q_i$.
\end{enumerate}
Relevance judgments can be obtained via human annotations or automatic evaluation metrics (e.g., BLEU, ROUGE, METEOR).

This evaluation framework enables the systematic benchmarking of LLM ranking models by providing a structured methodology for expertise identification and selection. While the framework is simplified for ranking-focused evaluation, it establishes a foundation for future extensions where expertise learning and answer generation can be jointly optimized. By constructing a reusable test collection, we facilitate the development of scalable, adaptable ranking methodologies for multi-LLM retrieval systems.

\subsection{Simulating Thousands of Expert LLMs}
\label{sec:simulation}

\begin{figure}[t]
        \centering
        \includegraphics[width=1\linewidth]{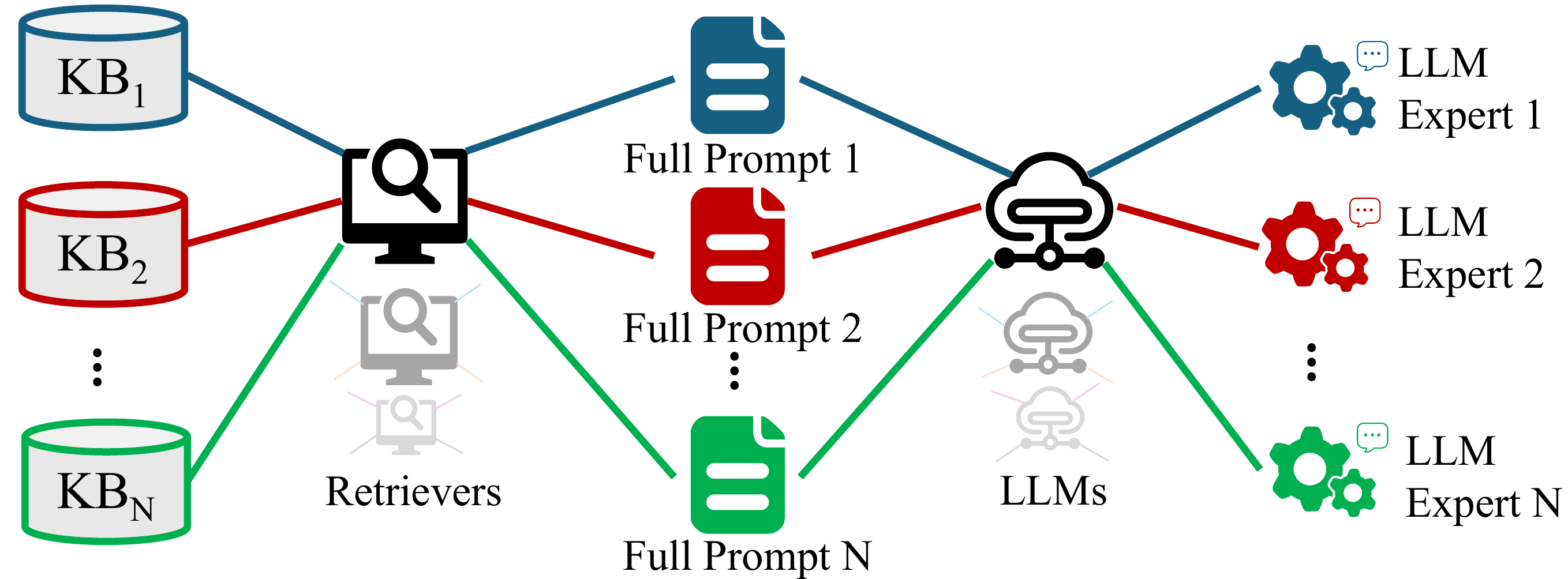}
        \caption{Using RAG-based LLMs to obtain expert models. Each expert consists of a shared LLM and retriever but is restricted to a distinct document collection, specifically structured KBs,  which define its domain expertise.}
        \label{fig:simulation}
\end{figure}

As previously mentioned, we do not provide direct access to LLMs but rather their responses. This raises the question: how can we obtain responses from thousands of domain-specific LLMs?
In practice, such a collection of LLMs, covering a wide variety of domains such as sports, finance, medicine, music, and cinema, does not yet exist, and it must be developed by us. We simulate the setup by leveraging a RAG-based approach, where each simulated expert consists of:  
\begin{itemize}
    \item A (shared) underlying LLM, responsible for generating responses. Note that instead of a single LLM it is possible to employ a small number of different LLMs, to diversify the language in the output. This is to address that sometimes, when only one LLM sits at the bottleneck, responses start to look templated in terms of writing style, and one may want to avoid that.
    \item A dedicated retriever, responsible for fetching relevant documents. Similarly to the previous point, one may also want to diversify the retrievers according to domain.
    \item A unique document collection, defining the domain expertise of the LLM.
\end{itemize}
Thus, an LLM configured with a cinema-related document collection acts as a cinema expert, while another with access to financial reports functions as a finance expert. 
This document-controlled retrieval ensures that each simulated expert has access only to domain-specific knowledge, allowing us to model the behavior of real-world expert LLMs. Our simulated setup is illustrated in Figure~\ref{fig:simulation}.

One of the key challenges in simulating expert LLMs is ensuring that they do not leak general knowledge from their base models. Ideally, an LLM should generate responses exclusively based on the retrieved documents and not rely on information it has learned during its pre-training phase. In practice, however, it is impossible to separate an LLM's language generation skills from its parametric memory.
To mitigate this issue, several strategies can be employed. In this work, we start simple and recommend (i) prompt engineering and (ii) query filtering. With prompt engineering, the model can be explicitly instructed to generate responses only based on the provided retrieved content. However, research has shown that this method is not always reliable, as LLMs sometimes disregard instructions and incorporate pre-trained knowledge into their responses. Query filtering is a more robust approach by which we filter user queries to ensure that the correct answers are absent from the LLM's pre-existing knowledge. This guarantees that any response must be derived purely from retrieval. One way to implement this is to compare base LLM responses against ground truth answers. If the base model fails to generate the correct answer without retrieval, but relevant documents are available, then we can further analyze whether the model's retrieved response logically follows (entails) the relevant documents. Queries in which the model's generated response is not entailed by the documents or differs significantly from the ground truth are ideal candidates for testing.

While an ideal evaluation setup would involve constructing a dedicated test collection with thousands of specialized document collections corresponding to different expert domains, this is computationally expensive and impractical at scale. Instead, a more feasible alternative is to reuse existing large-scale document collections.\looseness=-1

For this approach to work, the chosen collection must meet the following criteria:
\begin{itemize}
    \item It should be large enough to allow the construction of sub-collections representing different areas of expertise.
    \item It should include a large set of queries, for which either relevant documents are identified or ground truth answers have been collected.
\end{itemize}
Given such a dataset, a document clustering approach can be used to partition the full collection into thousands of clusters, each representing a distinct area of expertise.

For each simulated expert model, retrieval is performed only within its assigned cluster to obtain relevant documents. The retrieved documents are then used as input prompts to generate responses, ensuring that each expert model is constrained to its predefined domain. This method, however, presents a computational bottleneck, as it requires building thousands of individual search indices, one for each cluster. A more efficient alternative is to construct a single global index for the entire document collection. Then, upon receiving a query:
\begin{enumerate}
    \item Retrieve the top-k documents using the global index.
    \item Identify which clusters these documents belong to.
    \item For each cluster represented in the top-k ranking, provide the top-c documents to the LLM for response generation.
    \item For clusters that are not represented in the top-k ranking, a random document from the cluster can be sampled, since it is likely that all documents in the cluster are irrelevant to the query.
\end{enumerate}
This approach maintains domain specialization while significantly reducing computational complexity, making it a practical solution for scaling the simulation of thousands of expert LLMs.

\subsubsection{Implementation Details}

\begin{figure}[t!]
\begin{subfigure}{0.49\columnwidth}
  \centering
  \resizebox{1\columnwidth}{!}{\includegraphics[width=1.\columnwidth]{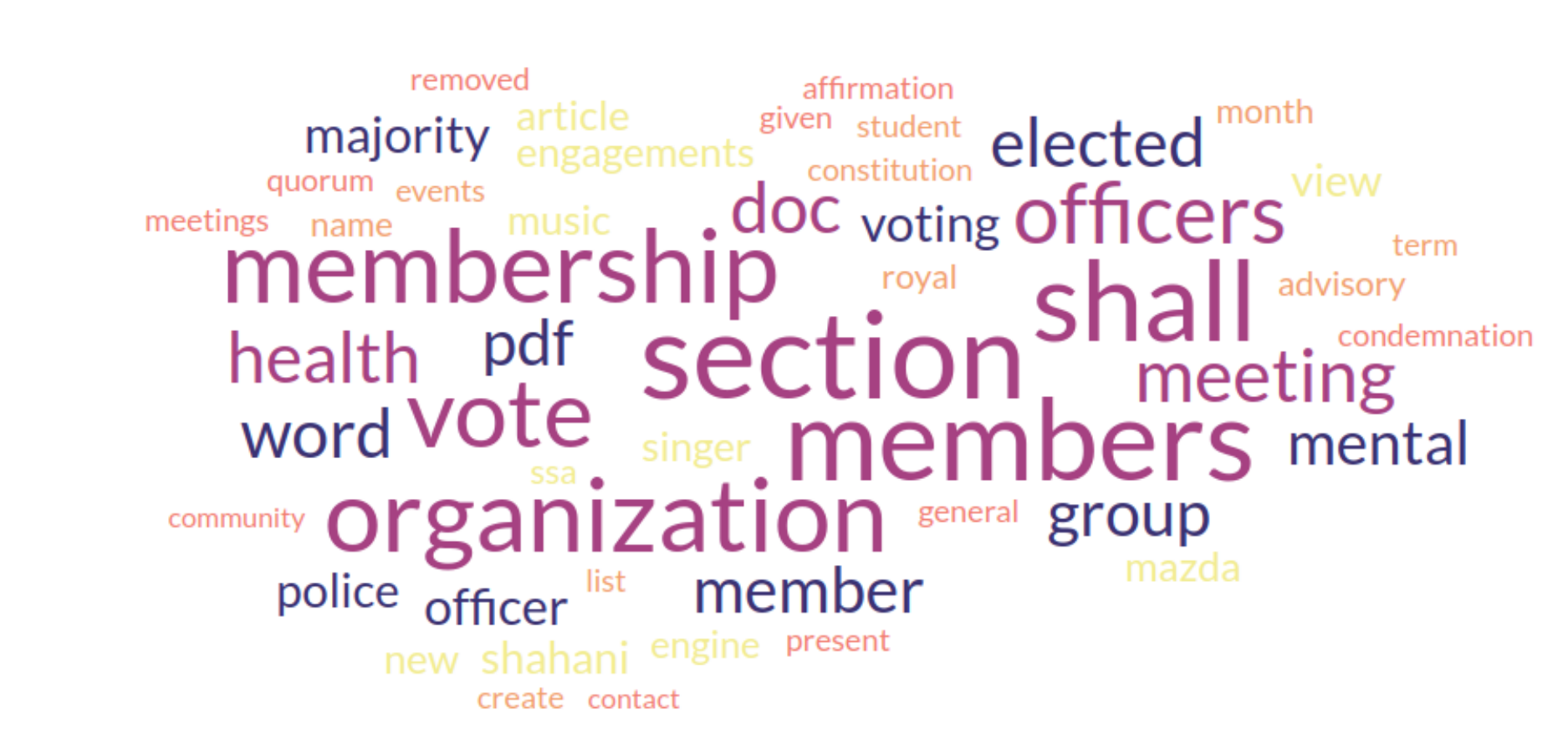}}
  \caption{Expertise in elections.}
  \label{fig:cluster_1}
\end{subfigure}
\begin{subfigure}{0.49\columnwidth}
  \centering
  \resizebox{1\columnwidth}{!}{\includegraphics[width=1.\columnwidth]{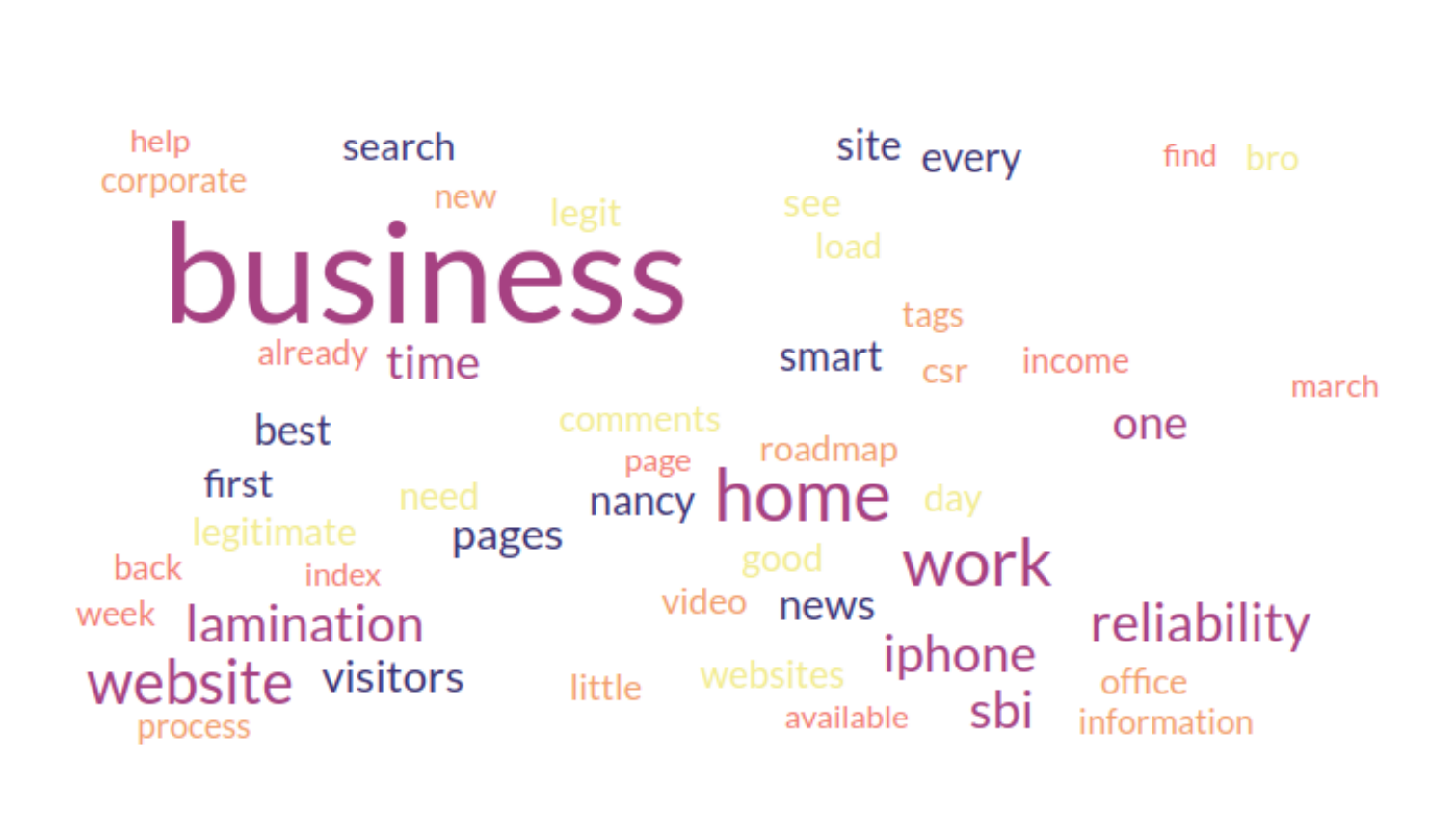}}
  \caption{Expertise in business.}
  \label{fig:cluster_2}
\end{subfigure}
\begin{subfigure}{0.49\columnwidth}
  \centering
  \resizebox{1\columnwidth}{!}{\includegraphics[width=1.\columnwidth]{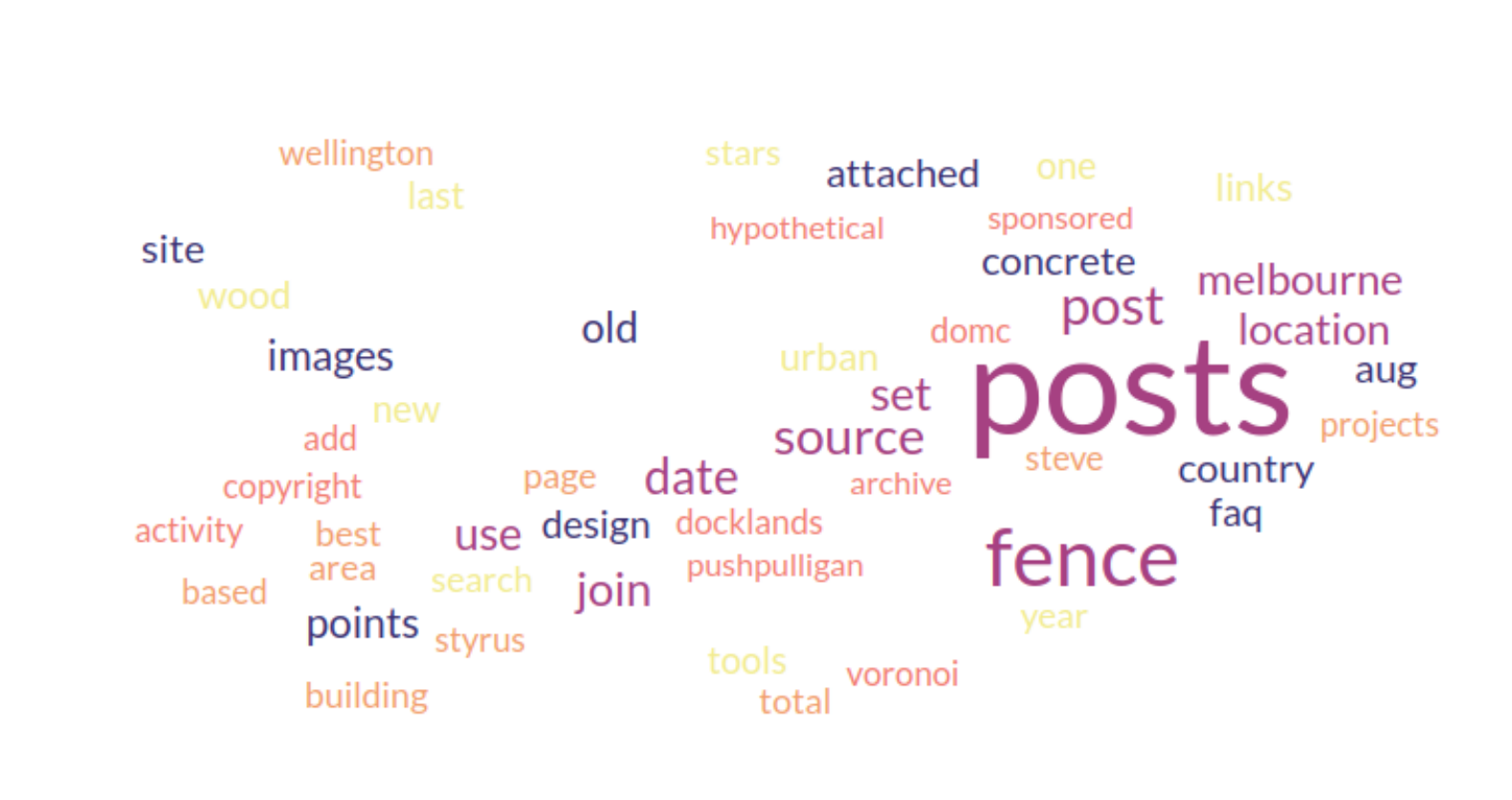}}
  \caption{Expertise in urban planning.}
  \label{fig:cluster_3}
\end{subfigure}
\begin{subfigure}{.49\columnwidth}
  \centering
  \resizebox{1\columnwidth}{!}{\includegraphics[width=1.\columnwidth]{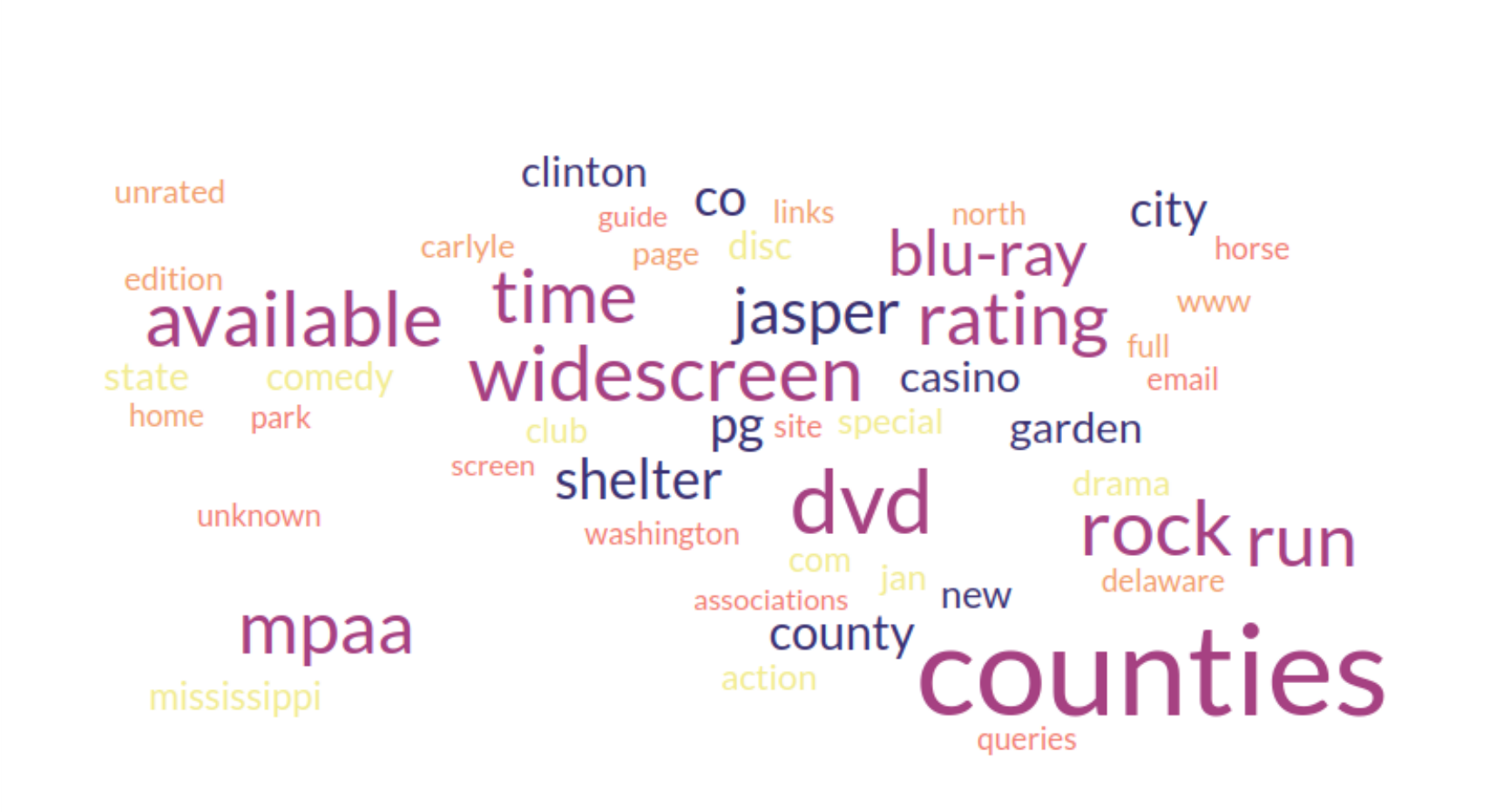}}
  \caption{Expertise in video.}
  \label{fig:cluster_4}
\end{subfigure}
\caption{Word clouds of 4 randomly chosen clusters, each simulating an LLM expert.}
\label{fig:word_cloud}
\end{figure}

\paragraph{Clustering} For effective creation of areas of expertise, a large document collection is necessary to ensure that the cardinality of each cluster is large. Existing collections that meet our standards are ClueWeb09~\cite{ClueWeb09}, containing approximately 504 million web pages, and Clueweb22~\cite{overwijk2022clueweb2210billionweb}, containing roughly 1 billion. 

To simulate the experts on ClueWeb09, we cluster its corpus with K-Means~\cite{Hartigan1979}, which ensures disjoint clusters with no overlap. Additionally, to transform the raw text to feature vectors, an efficient and scalable vectorizer with very low memory usage and streaming capability is essential for fast addition of new experts and practical constraints such as memory consumption. Instead of using a TF-IDF~\cite{SALTON1988513,7754750} vectorizer requiring in-memory vocabulary and mapping which renders it infeasible for very large collections on a resource-constrained infrastructure, a hashing vectorizer~\cite{Ramos1999,NIPS2017_f0f6ba4b,engproc2023046005} is used. A hashing vectorizer maps each text document to a fixed sized feature vector by applying a hash function to the term frequency of each token. The hashing vectorizer is stateless and can be easily used for addition of new expert knowledge. 1000-dimensional hash vectors of the collection were used for creating 2000 clusters for our document collections. Additionally, clusters with too many (generic) or too few (specific) documents were discarded. In Figure ~\ref{fig:word_cloud} we present a method by which we qualitatively estimate the expertise of a cluster; we sample random documents and assess their expertise based on their vocabulary. While this qualitative approach is rough, it makes for efficient human evaluation, given the scale of the problem.

Even though the Clueweb22 corpus is only twice as large as ClueWeb09, performing the same type of clustering using K-means and hashing on a billion documents is computationally prohibitive. Thankfully, the creators of Clueweb22 have tagged its web pages with topics. Leveraging this, we form approximately 20000 experts by performing topic-based clustering with a simple deterministic union-find algorithm.\looseness=-1

\paragraph{Question Answering} An essential component while building a data collection described in Section \ref{sec:test_collection} and the simulation process described in Section \ref{sec:simulation} is to ensure that ground truth answers are available for correctness and reliability of the question and answer pairs. To do so, we utilize QUASAR-T~\cite{dhingra2017quasardatasetsquestionanswering} as the annotated question answering dataset which is built over ClueWeb09 document collection. QUASAR-T contains a training set with $37012$ question-answer pairs, a dev set with $3139$ and a test set with $3000$. Each context document associated with a question-answer pair belongs to one of the clusters built in the clustering step, and is also associated with the selected cluster. Next, we built a global index using all the documents in the collection. Alternatively, $K$ indices can be built with the positive/relevant documents from Quasar-T and negative documents (which is not associated with a question-answer pair) from the same cluster. Each expert LLM uses the top-k relevant documents retrieved from the global/local index to simulate an expert by means of RAG.

\subsection{Discussion}

Our proposed evaluation framework has its own set of caveats. For instance, how does one create and query $K$ indices, when $K$ is large? What is the trade-off between having a single shared index between $K$ experts, and $K$ separate indices? However, dealing with these caveats is far more preferable to the alternative of storing and querying $K$ LLMs. For this reason, we believe that the simulation proposed in this section is a viable start to evaluating the research challenges proposed in Section~\ref{sec:challenges}.

\section{Conclusion}

The emergence of multi-LLM retrieval marks a significant departure from traditional information retrieval, shifting from document ranking to expert model selection and response synthesis. This evolution challenges long-established principles of retrieval effectiveness, and evaluation, demanding new methodologies that balance cost and accuracy. As we envision a future with millions of specialized LLMs, the necessity for scalable ranking mechanisms, robust expertise identification, and adaptive retrieval strategies becomes paramount.

The shift to multi-LLM retrieval presents several fundamental challenges. First, LLM expertise identification is complex, as models vary in domain specialization, reliability, and transparency, requiring new ranking mechanisms. Second, query efficiency and cost management are critical, as querying multiple LLMs is computationally expensive, necessitating budget-aware strategies that balance retrieval effectiveness with cost constraints. Third, multi-LLM response aggregation introduces challenges in synthesizing coherent, accurate, and unbiased answers from multiple models, especially when responses conflict. Fourth, trust, bias, and adversarial manipulation remain major concerns. Finally, evaluation frameworks must evolve, as traditional test collections and document relevance judgments are no longer sufficient. Our proposed evaluation framework lays the groundwork for systematically addressing these questions, yet it is only the beginning. The methodologies outlined here will require continuous refinement, driven by empirical research and large-scale experimental validation. 


\bibliographystyle{ACM-Reference-Format}
\bibliography{references}


\end{document}